\documentstyle[preprint,aps,epsfig,amsbsy,floats]{revtex}
 \tightenlines

\begin{document}
\newcommand{\beq}{\begin{equation}}
\newcommand{\eeq}{\end{equation}}
\newcommand{\beqa}{\begin{eqnarray}}
\newcommand{\eeqa}{\end{eqnarray}}
\newcommand{\sr}{\sqrt}
\newcommand{\fr}{\frac}
\newcommand{\mn}{\mu \nu}
\newcommand{\G}{\Gamma}

\draft \preprint{hep-th/0309179,~ INJE-TP-03-09}
\title{Relationship between five-dimensional
black holes and de Sitter spaces}
\author{  Y.S. Myung\footnote{E-mail address:
ysmyung@physics.inje.ac.kr}}
\address{
Relativity Research Center and School of Computer Aided Science\\
Inje University, Gimhae 621-749, Korea} \maketitle

\begin{abstract} We study a close relationship between  the topological
anti-de Sitter (TAdS)-black holes and topological de Sitter (TdS)
spaces including the Schwarzschild-de Sitter (SdS) black hole in
five-dimensions. We show that all thermal properties of the TdS
spaces can be found from those of the TAdS black holes by
replacing $k$ by $-k$. Also we find that all thermal information
for the cosmological horizon of the SdS black hole is obtained
from either the hyperbolic-AdS black hole or the Schwarzschild-TdS
space by substituting $m$ with  $-m$. For this purpose we
calculate thermal quantities of bulk, (Euclidean) conformal field
theory (ECFT) and moving domain wall  by using the A(dS)/(E)CFT
correspondences. Further we compute logarithmic corrections to the
Bekenstein-Hawking entropy, Cardy-Verlinde
 formula and Friedmann equation  due to
 thermal fluctuations. It implies that the cosmological horizon
of the TdS spaces is nothing but the event horizon of the TAdS
black holes and the dS/ECFT correspondence is valid for the TdS
spaces in conjunction with the AdS/CFT correspondence for the TAdS
black holes.

\end{abstract}

\newpage
\section{Introduction}
Recently an accelerating universe has proposed to be a way to
interpret the astronomical data of supernova\cite{Per,CDS,Gar}.
The inflation has been employed to solve the cosmological flatness
and horizon puzzles arisen in the standard cosmology. Combining
the accelerating universe with the need of inflation
 leads to that our universe approaches de Sitter
geometries in both the infinite past and the infinite
future\cite{Wit,HKS,FKMP}. Hence it is  important to study the
nature of de Sitter (dS) space and the assumed dS/CFT
correspondence\cite{BOU,TDS,STR}.
 However,
some difficulties appeared in studying de Sitter space with a
positive cosmological constant $\Lambda_{dS}>0$. i) There is no
spatial (timelike) infinity and global timelike Killing vector.
Thus it is not easy to define conserved quantities including mass,
charge and angular momentum appeared in asymptotically  dS space.
ii) The dS solution is absent from string theories and thus we do
not find  a definite example to test the assumed dS/CFT
correspondence. iii) It is hard to define  the $S$-matrix because
of the presence of the cosmological horizon\footnote{In anti de
Sitter space (AdS) with a negative cosmological constant
$\Lambda_{AdS}<0$, all of three difficulties mentioned in dS space
seem to be resolved even though lacking for a globally defined
timelike Killing vector\cite{Wit}. There exists a spatial
(timelike) infinity and hence the region outside the event horizon
is noncompact. There are many AdS solutions arisen from string
theory or $M$-theory. Even though there is no notion of an
$S$-matrix in asymptotically AdS space,  one has the correlation
functions of the boundary CFT. That is, the correlators of the
boundary CFT may provide the $S$-matrix elements of the $\Lambda
\to 0$ theory.}.

On the other hand, we remind the reader that the cosmological
horizon in dS space is very similar to the event horizon in the
sense that one can define its thermodynamic quantities of a
temperature and an entropy using the same way as was done for the
 black hole\cite{GHaw}.  We note that the
five-dimensional Schwarzschild black hole which is asymptotically
flat has a negative specific heat of $C_v^{Sch}=-3S_0$ with
 the Bekenstein-Hawking entropy ($S_0$)~\cite{DMB}.
 This means that the Schwarzschild black hole is never
in thermal equilibrium and it evaporates according to the Hawking
radiation. But the Schwarzschild black hole could be thermal
equilibrium with a radiation in a bounded box.  This is because
the black hole has a negative specific heat while the radiation
has a positive one. The two will be in thermal equilibrium if the
box is bounded. The Schwarzschild-AdS black hole  belongs to this
category  and its specific heat takes the form of $C_v^{SAdS}
\simeq 3S_0$ for large black hole. Also we note that a
cosmological horizon in five-dimensional de Sitter space has a
positive specific heat of $C_v^{dS}=3S_0$. This implies that the
cosmological horizon in de Sitter space is  rather similar to the
Schwarzschild-AdS black hole  than the Schwarzschild black hole.
Furthermore, for large black holes including the Schwarzschild-AdS
black hole, the Bekenstein-Hawking entropy receives logarithmic
corrections due to thermodynamic fluctuations~\cite{KM1}.

In this work, we establish a close relationship between  the
topological anti-de Sitter (TAdS)-black holes and topological de
Sitter (TdS) spaces including the Schwarzschild-de Sitter (SdS)
black hole in five-dimensions. We show that all thermal properties
of the TdS spaces can be found from those of the TAdS black holes
by replacing $k$ by $-k$. Also we find that all thermal
information for the cosmological horizon of the SdS black hole is
obtained from either the hyperbolic-AdS black hole or the
Schwarzschild-TdS space by substituting $m$ with  $-m$. For this
purpose we calculate thermal quantities of bulk, (Euclidean)
conformal field theory  and moving domain wall  by using the
A(dS)/(E)CFT correspondences. Further we compute logarithmic
corrections to the Bekenstein-Hawking entropy, Cardy-Verlinde
 formula and Friedmann equation due to
 thermal fluctuations. We conclude that the cosmological horizon
of the TdS spaces is nothing but the event horizon of the TAdS
black holes and the dS/ECFT correspondence is valid for the TdS
spaces in conjunction with the AdS/CFT correspondence for the TAdS
black holes. However, the dS/ECFT correspondence is not clearly
realized in the SdS black hole. Hence the SdS black hole does not
seems to be a toy model to study the cosmological horizon in
asymptotically de Sitter space.

The organization of this paper is as follows. In section II we
briefly review the bulk thermal property. Section III is devoted
to studying their boundary thermal CFT.  We obtain the same form
of the  Cardy-Verlinde formula.  In order to establish  the
dynamical A(dS)/(E)CFT correspondence, we study the moving domain
wall approach in section IV. In section V we obtain logarithmic
corrections to the Cardy-Verlinde formula. Using the  holographic
principle, in section VI we find the modified Friedmann equations
including the logarithmic terms. Finally we discuss our results in
section VII.

\section{Bulk Thermal Property}
\subsection{Topological AdS black holes}
 The topological AdS black holes
in five-dimensional spacetimes are given by~\cite{BIR}
 \beq ds^{2}_{TAdS}=
 -h_{TAdS}(r)dt^2 +\fr{1}{h_{TAdS}(r)}dr^2 +r^2d\Sigma_k^2,~~d\Sigma_k^2= \left[d\chi^2
+f_{k}(\chi)^2(d\theta^2+ \sin^2 \theta d\phi^2) \right],
\label{BMT} \eeq where $d\Sigma_k^2=\gamma^k_{ij}dx^idx^j$
describes the horizon geometry with a constant curvature of $6k$.
$h_{TAdS}(r)$ and $f_k(\chi)$ are given by \beq
h_{TAdS}(r)=k-\fr{m}{r^2}+ \fr{ r^2}{\ell^2},~~~ f_{0}(\chi)
=\chi, ~f_{1}(\chi) =\sin \chi, ~f_{-1}(\chi) =\sinh \chi. \eeq
Here we define $k$=1,~0, and $-1$ cases as the Schwarzschild-AdS
(SAdS) black hole~\cite{MP}, flat-AdS (FAdS) black hole, and
hyperbolic-AdS (HAdS) black hole~\cite{CAI1}, respectively.
Choosing $m>0$, the only event horizon is given by
 \beq \label{EH} r_{EH}^2=
\fr{\ell^2}{2}\Big(-k+\sqrt{k^2 +4 m/\ell^2}\Big). \eeq For $k=1$,
we have both a small black hole ($r_{EH}^2<<\ell^2,4m<<\ell^2$)
 with  the horizon at $r=r_{EH}$, where $r_{EH}^2
 \simeq m$ and a large black hole  ($r_{EH}^2>>\ell^2,4m>>\ell^2 )$  with  the
horizon at $r=r_{EH}$ given by $r_{EH}^2 \simeq \sqrt{m}\ell$. For
$k=0$ case, one has the event horizon  at $r=r_{EH}$, where
$r_{EH}^2 = \sqrt{m}\ell$. In the case of $k=-1$, for $4m<<\ell^2$
one has the event horizon at $r=r_{EH}$, where $r_{EH}^2 \simeq
\ell^2+ m$ and for $4m>>\ell^2$ one has the  event horizon at
$r=r_{EH}$ given by $r_{EH}^2 \simeq \sqrt{m}\ell$. That is, one
always finds $r_{EH}^2>\ell^2$ for $k=-1$\footnote{We note that
for $k=-1,~m<0$, the event horizon of this HAdS black hole leads
to the cosmological horizon of the Schwarzschild-de Sitter black
hole Eq.(\ref{3EH}).}.

The relevant thermodynamic quantities: reduced mass ($m$), free
energy ($F$), Bekenstein-Hawking entropy ($S_0$), Hawking
temperature ($T_H$),  energy (ADM mass :$E=M$), and specific heat
($C_v=(dE/dT)_V$) are given by~\cite{LNOO} \beqa \label{TQ}
&&m=r_{EH}^2 \Big(\fr{r_{EH}^2}{\ell^2}+ k \Big),~~F=-\fr{V_3
r_{EH}^2}{16 \pi G_5}\Big(\fr{r_{EH}^2}{\ell^2}- k \Big),~~
S_0=\fr{V_3r_{EH}^3}{4G_5},~~\\ \nonumber &&T_H=\fr{k}{2 \pi
r_{EH}} +\fr{r_{EH}}{\pi \ell^2}~~E=F+T_H S_0 =\fr{3V_3m}{16 \pi
G_5}=M,~~C_v= 3 \fr{2r_{EH}^2+k\ell^2}{2r_{EH}^2-k\ell^2}S_0 \eeqa
where  $G_5$ is the five-dimensional Newton constant. Here
$V_{3(k)}=\int dx^3 \sqrt{\gamma^k}$ is the volume of unit
three-dimensional hypersurface. As an example, $V_{3(k=1)}=2\pi^2$
is the volume of a unit $S^3$. For simplicity we use an implicit
notation of $V_3$ instead of $V_{3(k)}$ otherwise state.  All
thermodynamic quantities except for $F$ and $C_v$ are positive for
any $k$. In the limit of $\ell\to \infty$, we recover the negative
specific heat ($C_v^{Sch}=-3S_s$) of the Schwarzschild black hole.
On the other hand, in the limit of $\ell\to 0$  one finds a
positive value of $C_v^{\ell \to 0}=3S_0$ for the large SAdS-black
hole.

\subsection{Schwarzschild de Sitter black hole} In order to find
the thermal property of a black hole in asymptotically de Sitter
space, we consider Schwarzschild de Sitter (SdS) black hole in
five-dimensional spacetimes~\cite{CAI2}  \beq ds^{2}_{SdS}=
-h_{SdS}(r)dt^2 +\fr{1}{h_{SdS}(r)}dr^2 +r^2 d\Sigma_{k=1}^2
\label{SDS} \eeq where $h_{SdS}(r)$ is given by \beq
h_{SdS}(r)=1-\fr{m}{r^2}- \fr{ r^2}{\ell^2}.\eeq
 In
the case of $m=0$, we have an exact de Sitter space with its
curvature radius $\ell$. However, $m>0$ generates the SdS black
hole.  Here we have  two horizons.   The cosmological and event
horizons are given by \beq \label{3EH} r_{CH/EH}^2=
\fr{\ell^2}{2}\Big(1\pm \sqrt{1 -4 m/\ell^2}\Big). \eeq We
classify three cases : 1) $4m=\ell^2$, 2) $4m>\ell^2$, 3)
$4m<\ell^2$. The case of $4m=\ell^2$ corresponds to the maximum
black hole and the minimum cosmological horizon in asymptotically
de Sitter space (that is, Nariai black hole). In this case we have
$r_{EH}^2=r_{CH}^2=\ell^2/2=2m$. The case of $4m>\ell^2$ is not
allowed for the black hole in de Sitter space. The case of
$4m<\ell^2$ corresponds to a small black hole within the
cosmological horizon. In this case we have the cosmological
horizon at $r=r_{CH}$, where $r_{CH}^2 \simeq \ell^2-m$ and the
event horizon at $r=r_{EH}$ given by $r_{EH}^2 \simeq m$. Hence we
find  two important relations for the SdS solution: \beq
\label{2Ineq} m \le r^2_{EH} \le \ell^2/2,~~ \ell^2/2 \le r^2_{CH}
\le \ell^2-m \eeq which means that as $m$ increases from a small
value to the maximum  of $m=\ell^2/4$, a small black hole
increases up to the Nariai black hole. On the other hand the
cosmological horizon decreases from a large one of $(\ell^2-m)$ to
the minimum of $\ell^2/2$.

 The thermodynamic quantities for two horizons   are given
by~\cite{CM,NOO} \beqa \label{2TQ} &&m=r_{EH/CH}^2
\Big(-\fr{r_{EH/CH}^2}{\ell^2}+ 1 \Big),~~F_{EH/CH}=\pm\fr{V_3
r_{EH/CH}^2}{16 \pi G_5}\Big(\fr{r_{EH}^2}{\ell^2}+1 \Big),~~
S_0=\fr{V_3r_{EH/CH}^3}{4G_5},~~\\ \nonumber
&&T_H^{EH/CH}=\pm\fr{1}{2 \pi r_{EH/CH}} \mp \fr{r_{EH/CH}}{\pi
\ell^2},~~E=F_{EH/CH}+T_H^{EH/CH} S_0 =\pm \fr{3V_3m}{16 \pi
G_5},\\ \nonumber &&C_v^{EH/CH}= 3
\fr{2r_{EH/CH}^2-\ell^2}{2r_{EH/CH}^2+\ell^2}S_0,
 \eeqa where $V_3$ denotes the volume of a unit
three-dimensional sphere : $V_{3(k=1)}$.  In the limit of $\ell\to
\infty$, we recover the negative specific heat ($C_v^{Sch}=-3S_s$)
of the Schwarzschild black hole. Considering Eqs.(\ref{2Ineq}) and
(\ref{2TQ}), all quantities except for $F_{CH},~E_{CH},~C_v^{EH}$
belong positive.  On the other hand, in the limit of $\ell\to 0$
one finds a positive value of $C_v^{dS}=3S_0$ for the exact de
Sitter space.

\subsection{Topological de Sitter space} The topological de Sitter
(TdS) solution was originally introduced to check the mass bound
conjecture in de Sitter space: any asymptotically de Sitter space
with the mass greater than exact de Sitter space has a
cosmological singularity~\cite{TDS}.  For our purpose, we consider
the topological de Sitter solution in five-dimensional spacetimes
\beq ds^{2}_{TdS}= -h_{TdS}(r)dt^2 +\fr{1}{h_{TdS}(r)}dr^2 +r^2
d\Sigma_k^2, \label{Tds} \eeq where  $h_{TdS}(r)$ is given by \beq
h_{TdS}(r)=k+\fr{m}{r^2}- \fr{ r^2}{\ell^2}. \eeq Requiring $m>0$,
the black hole disappears and instead a naked singularity occurs
at $r=0$. Here we define $k=1,0,-1$ cases as the
Schwarzschild-topological de Sitter (STdS) space, flat-topological
de Sitter (FTdS) space, and hyperbolic--topological de Sitter
(HTdS) space. In the case of $k=1,m=0$, we have an exact de Sitter
space with its curvature radius $\ell$. However, $m>0$ generates
the topological de Sitter spaces. The only cosmological horizon
exists as \beq \label{CH} r_{CH}^2= \fr{\ell^2}{2}\Big(k+\sqrt{k^2
+4 m/\ell^2} \Big). \eeq

 For $k=-1$ case we have both a small cosmological horizon ($r_{CH}^2<<\ell^2,4m<<\ell^2$)
 with  the horizon at $r=r_{CH}$, where $r_{CH}^2
 \simeq m$ and a large cosmological horizon ($r_{CH}^2>>\ell^2,4m>>\ell^2 )$  with  the
horizon at $r=r_{CH}$ given by $r_{CH}^2 \simeq \sqrt{m}\ell$. For
$k=0$ case, one has the cosmological horizon  at $r=r_{CH}$, where
$r_{CH}^2 = \sqrt{m}\ell$. In the case of $k=1$, for $4m<<\ell^2$
one has the  cosmological horizon at $r=r_{CH}$, where $r_{CH}^2
\simeq \ell^2+ m$ and for $4m>>\ell^2$, one has the cosmological
horizon at $r=r_{CH}$, where $r_{CH}^2 \simeq \sqrt{m}\ell$. Here
we have $r_{CH}^2>\ell^2$ for $k=1$ case\footnote{For $k=1,m<0$,
the cosmological horizon of the STdS space leads to the
cosmological horizon of the SdS black hole spacetime
Eq.(\ref{3EH}).}.

The  thermodynamic quantities for the cosmological horizon are
calculated as~\cite{CAI2,myung}\beqa \label{3TQ} &&m=r_{CH}^2
\Big(\fr{r_{CH}^2}{\ell^2}- k \Big),~~F=-\fr{V_3 r_{CH}^2}{16 \pi
G_5}\Big(\fr{r_{CH}^2}{\ell^2}+ k \Big),~~
S_0=\fr{V_3r_{CH}^3}{4G_5},~~\\ \nonumber &&T_H=-\fr{k}{2 \pi
r_{CH}} +\fr{r_{CH}}{\pi \ell^2},~~E=F+T_HS =\fr{3V_3m}{16 \pi
G_5}=M,~~C_v= 3 \fr{2r_{CH}^2-k\ell^2}{2r_{CH}^2+k\ell^2}S_0,
\eeqa where $V_3$ is the volume of a unit three-dimensional
hypersurface :$V_{3(k)}$. All thermal quantities except for $F$
and $C_v$ are positive ones. It is found from Eqs.(\ref{TQ}),
(\ref{2TQ}) and (\ref{3TQ}) that all thermodynamic results of the
TdS solution can be recovered from those of the TAdS solution by
replacing $k$ by $-k$. Also we find that all thermal information
for the cosmological horizon of the SdS black hole is obtained
from either the HAdS black hole or the STdS space by substituting
$m$ with $-m$. However  we note that for the same $m,\ell$,
thermal quantities of the TdS case with $k$ are not precisely
equal to those of the TAdS case with $-k$ because of
$V_{3(k=\pm1)}\not=V_{3(k=\mp1)}$. Also thermal quantities of the
CSdS with $V_{3(k=1)}$ is not exactly the same as in the HAdS case
with $V_{3(k=-1)}$. In this sense we use a notation of $\simeq$ in
TABLE.

\section{Boundary CFT and Cardy-Verlinde formula}
\begin{table}
 \caption{Summary of bulk specific heats, boundary (E)CFT energy and  Casimir energy
 for 5D TAdS black holes, TdS spaces and SdS black hole. In the moving domain wall approach, the energy
 density term in Friedmann equation is included for comparison with the boundary (E)CFT energy. }
 \begin{tabular}{lp{4.5cm}p{3cm}}
 thermodynamical system   & $C_v$ & $E_4/\rho_{(E)CFT}/E_c $\\ \hline
  HAdS & + & +/+/$-$ \\
FAdS & +($3S_0$) & +/+/0 \\
SAdS & + if $r^2_{EH}>
\ell^2/2$ & +/+/+ \\
STdS($\simeq$ HAdS) & + & +/+/$-$ \\
FTdS($\simeq$ FAdS) & + ($3S_0$) & +/+/$0$ \\
HTdS($\simeq$ SAdS) & + if $ r^2_{CH}>
\ell^2/2$& +/+/+ \\
ESdS & $-$ & +/$-$/+ \\
CSdS($\simeq$ HAdS,~STdS if $ m \to -m$)  & + & $-/-/-$
 \end{tabular}
 \end{table}
The holographic principle means that the number of degrees of
freedom associated with the bulk gravitational dynamics is
determined by its boundary spacetime. The AdS/CFT correspondence
represents a realization of this principle~\cite{HOL}. For a
strongly coupled CFT with its AdS dual, one obtains the
Cardy-Verlinde formula~\cite{VER}. Indeed this formula holds for
various kinds of asymptotically AdS spacetimes including the TAdS
black holes~\cite{CAI1}. Also it holds for asymptotically de
Sitter spacetimes including the SdS black hole and TdS
spaces~\cite{CAI2}.  The boundary spacetimes for the (E)CFT are
defined through the A(dS)/CFT correspondences~\cite{witten} \beqa
\label{bcft} &&ds^2_{CFT}=\lim_{r \to
\infty}\fr{R^2}{r^2}ds^2_{TAdS}=-\fr{R^2}{\ell^2}dt^2 +R^2
d\Sigma^2_k,\\
&& ds^2_{ECFT}=\lim_{r \to \infty}
\fr{R^2}{r^2}ds^2_{SdS}=\fr{R^2}{\ell^2}dt^2 +R^2
d\Sigma^2_{k=1},\\
&&ds^2_{ECFT}=\lim_{r \to \infty}\fr{R^2}{r^2}
ds^2_{TdS}=\fr{R^2}{\ell^2} dt^2 +R^2 d\Sigma^2_k. \eeqa From the
above,  the relation between the five-dimensional bulk and
four-dimensional boundary quantities is given by $E_4=(\ell/R)E,~
T=(\ell/R)T_H$ where $R$ satisfies $T>1/R$ but one has the same
entropy : $S_4=S_0$.  We note that the boundary physics is
described by the CFT-radiation matter with the equation of state:
$p=E_4/3V_3$. Then  the Casimir energy is given by
$E_c=3(E_4+pV_3-TS_0)$. We obtain the boundary thermal quantities
\beqa &&E_4^{TAdS}=\fr{3V_3m\ell}{16 \pi G_5
R},~~T^{TAdS}=\fr{k\ell}{2\pi r_{EH} R}+ \fr{r_{EH}}{\pi \ell
R},~~E_c^{TAdS}=k \fr{3\ell r_{EH}^2 V_3}{8
\pi G_5 R},\\
 &&E_4^{ESdS/CSdS}=\pm \fr{3V_3m\ell}{16 \pi G_5
 R},~~T^{ESdS/CSdS}=\pm\fr{\ell}{2\pi r_{EH/CH} R}\mp \fr{r_{EH/CH}}{\pi \ell
R},\\ \nonumber &&E_{c}^{ESdS/CSdS}=\pm \fr{3\ell
r_{EH/CH}^2 V_3}{8 \pi G_5 R},\\
 &&E_4^{TdS}=\fr{3V_3m\ell}{16 \pi G_5 R},T^{TdS}=-\fr{k\ell}{2\pi r_{CH} R}+ \fr{r_{CH}}{\pi \ell
R},~~~~E_c^{TdS}=-k \fr{3\ell r_{CH}^2 V_3}{8 \pi G_5 R}, \eeqa
where ESdS (CSdS) represent   the event horizon (cosmological
horizon) of the SdS black hole. For the ESdS, the substitution
rule of $m \to -m$ is no longer valid for deriving
$E_4^{ESdS},E_c^{ESdS}$ from either the HAdS with $k=-1$ or the
STdS with $k=1$. Using this expression, one finds the
Cardy-Verlinde formula\footnote{For the topological
Reissner-Nordstrom-de Sitter (TRNdS) black hole, see
ref.\cite{SET}.}\beqa \label{cav1} && TAdS~ :~~S_{0CV}^{TAdS}=\fr{
2 \pi
R}{3 \sqrt{|k|}} \sqrt{|E_c(2E_4-E_c)|} , \\
\label{cav2}&& SdS~:~~S_{0CV}^{SdS}=\fr{ 2 \pi R}{3} \sqrt{|E_c(2E_4-E_c)|}, \\
\label{cav3}&& TdS~:~~S_{0CV}^{TdS}=\frac{ 2 \pi R}{3 \sqrt{|k|}}
\sqrt{|E_c(2E_4-E_c)|}.  \eeqa All static thermodynamic
information is encoded in TABLE. It is shown that all thermal
properties of the TdS spaces can be found from those of the TAdS
black holes by replacing $k$ by $-k$. Also we find that all
thermal information for the cosmological horizon of the SdS black
hole is obtained from either the event horizon of the HAdS black
hole or the cosmological horizon of the  STdS space by
substituting $m$ with $-m$. Concerning the A(dS)/CFT
correspondences, we remind the reader that the boundary CFT energy
($E_4$) should be positive in order for it to make sense. However,
one
 finds  that $E_4^{CSdS}<0$ for the cosmological horizon of the SdS black
 hole. It suggests that the dS/CFT correspondence is not valid for this case.
 Also the Casimir energy ($E_c$) is related to the central charge
 of the corresponding CFT. Hence if it is negative, one  may obtain
 a non-unitary CFT. In this sense, HAdS, STdS, and CSdS cases may be
 problematic. In order to understand the negative energy of $E_4^{CSdS}<0$,
 we have to study the dynamic A(dS)/(E)CFT correspondences in the next
 section.

\section{Moving domain wall(MDW) approach}
\subsection{MDW in TAdS black holes}
Now we introduce the radial  location of a MDW in the form of $
r=a(\tau),t=t(\tau)$ parameterized by the proper time $\tau$ :
$(t,r,\chi,\theta,\phi)\to(t(\tau),a(\tau), \chi, \theta, \phi)$.
Then we expect that the induced metric of moving  domain wall
(brane) will be given by the  FRW-type. Hence $\tau$ and $a(\tau)$
will imply the cosmic time and scale factor of the FRW-universe,
respectively. A tangent vector (proper velocity)  of this MDW

\beq u= \dot t \fr{\partial}{\partial t}+ \dot a
\fr{\partial}{\partial a},\label{TAN} \eeq is introduced to define
an embedding properly. Here  overdots mean  differentiation with
respect to $\tau$. This is normalized to  satisfy \beq u^{M}u^{N}
g_{MN}=-1. \label{NTA} \eeq Given a tangent vector $u_M$, we need
a normal 1-form
 directed toward to the bulk. Here we choose this as
\beq n= \dot a dt-  \dot t da, ~~~n_{M}n_{N} g^{MN}=1. \label{NNV}
\eeq

Using either Eq.(\ref{TAN}) with (\ref{NTA}) or Eq.(\ref{NNV}), we
can express the proper time rate of the TAdS time
 $\dot t$ in terms of $\dot a$
as \beq \dot t=\fr{\sqrt{\dot a^2 +h_{TAdS}(a)}}{h_{TAdS}(a)}.
\label{TAV} \eeq The first two terms in Eq.(\ref{BMT}) together
with Eq.(\ref{TAV}) leads to a timelike brane \beq
-h_{TAdS}(r)dt^2 + \fr{1}{h_{TAdS}(r)} dr^2 ~~\to~~
-\Big(h_{TAdS}(a) \dot t^2 -\fr{\dot a^2}{h_{TAdS}(a)}
\Big)d\tau^2= -d\tau^2. \eeq Then the 4D induced line element is
\beqa ds^{2}_{4TAdS}&&=-d \tau^2 +a(\tau)^2 d\Sigma^2_k
\nonumber \\
&&\equiv h_{\mu \nu}dx^{\mu} dx^{\nu}, \label{INM} \eeqa where we
use the Greek indices only  for the brane. Actually the embedding
of the FRW-universe  is a $2(t,r) \to 1(\tau)$-mapping. The
projection tensor is given by $h_{MN}=g_{MN}-n_Mn_N$ and its
determinant is zero. Hence its inverse  $h^{MN}$ cannot be
defined. This means that the above embedding belongs to a peculiar
mapping to obtain the induced metric $h_{\mu\nu}$ from the TAdS
black hole spacetime $g_{MN}$ together with $n_M$. In addition,
the extrinsic curvature is defined  by

\beqa &&K_{\tau\tau}=K_{MN} u^M u^N =(h_{TAdS}(a) \dot
t)^{-1}(\ddot a +h_{TAdS}'(a)
 /2)=\fr{\ddot a +h_{TAdS}'(a)/2}
{\sqrt{\dot a^2 +h_{TAdS}(a)}}, \\
&&K_{\chi\chi} = K_{\theta\theta}=K_{\phi\phi} =- h_{TAdS}(a) \dot
t a=-\sqrt{\dot a^2 +h_{TAdS}(a)}~a. \eeqa A localized matter on
the brane implies that the extrinsic curvature jumps across the
brane. This jump is described by the Israel junction condition

\beq K_{\mu \nu}=-\kappa^2 \left(
T_{\mu\nu}-\fr{1}{3}T^{\lambda}_{\lambda}h_{\mu\nu} \right)
\label{4DI} \eeq with $\kappa^2=8 \pi G_5$.  We   introduce a
localized stress-energy tensor on the brane as the 4D perfect
fluid

\beq T_{\mu \nu}=(\varrho +p)u_{\mu}u_{\nu}+p\:h_{\mu\nu}.
\label{MAT} \eeq Here $\varrho=\rho+ \sigma$ $(p=P-\sigma)$, where
$\rho $ $(P)$ is the energy density (pressure) of the localized
matter and $\sigma$ is the brane tension. In the case of
$\rho=P=0$, the r.h.s. of Eq.(\ref{4DI}) leads to  a form of the
Randall-Sundrum case as $-\fr{\sigma \kappa^2}{3} h_{\mu\nu}$.
From Eq.(\ref{4DI}), one finds
 the space component of the junction condition

\beq \sqrt{ \dot a^2+h_{TAdS}(a)}=\fr{\kappa^2}{3}\sigma a.
\label{SEE} \eeq For a single TAdS space, we introduce the
fine-tuned brane tension $\sigma=3/(\kappa^2\ell)$ to obtain a
critical brane. Then the above equation leads to \beq H^2_{TAdS}=-
\fr{k}{a^2} +\fr{m_{TAdS}}{a^4}, \label{HHH} \eeq where $H=\dot
a/a$ is the expansion rate and $m_{TAdS}$ is given in
Eq.(\ref{TQ}). The term of $m_{TAdS}/a^4$ originates from the
electric (Coulomb) part of the five-dimensional Weyl tensor,
$E_{00} \sim m/r^2$. This term behaves like a radiation\cite{SV}.
Especially for the SAdS, we have $m_{TAdS}=16 \pi G_5M_{TAdS}/3
V_3,~M_{TAdS}=a \tilde E_{TAdS}/\ell,
 V=a^3 V_3, G_5=\ell G_4/2$. Then one finds a CFT-radiation dominated
universe \beq H^2_{SAdS}=- \fr{1}{a^2} +\fr{8\pi
G_4}{3}\rho_{CFT},~~~\rho_{CFT}=\fr{\tilde E_{TAdS}}{V}.
\label{1CRA} \eeq The equation (\ref{SEE}) is well-defined even at
$a=r_{EH}$. Thus Eq.(\ref{SEE}) leads to $H^2_{TAdS}=1/\ell^2$
when the MDW crosses the event horizon of the TAdS black hole
spaces.

\subsection{MDW in SdS black hole}
As is shown in the  section III, to obtain a spacelike MDW with
Euclidean signature in the SdS background we have to use a
different mapping. The first two terms in Eq.(\ref{SDS}) together
with $u^Mu^Ng_{MN}=1;~n^Mn^Ng_{MN}=-1$ leads to a spacelike brane
\beq -h_{SdS}(r)dt^2 + \fr{1}{h_{SdS}(r)} dr^2 ~~\to~~
-\Big(h_{SdS}(a) \dot t^2 -\fr{\dot a^2}{h_{SdS}(a)} \Big)d\tau^2=
d\tau^2. \eeq Then the 4D induced line element is \beq
ds^{2}_{4SdS}=d \tau^2 +a(\tau)^2 d\Sigma^2_{k=1}. \label{INM}
\eeq The Israel junction condition leads to \beq \label{2ISR}
\sqrt{ \dot a^2-h_{SdS}(a)}=\fr{\kappa^2}{3}\sigma a.\eeq
 For a single SdS black hole, we choose the
brane tension $\sigma=3/(\kappa^2\ell)$ to obtain a critical
brane. Eq.(\ref{2ISR}) leads to \beq H^2_{SdS}=\fr{1}{a^2}
-\fr{m_{SdS}}{a^4}. \label{2HHH} \eeq Here we find a negative term
of $-m_{SdS}/a^4$ where $m_{SdS}$ is defined in Eq.(\ref{2TQ}).
This term behaves like an exotic radiation\cite{MYU2}. Further  we
have $m_{SdS}=16 \pi G_5M_{SdS}/3 V_3$ and $M_{SdS}=a \tilde
E_{SdS}/\ell$. Then one finds an exotic ECFT-radiation dominated
universe \beq \label{2CRA} H^2_{SdS}= \fr{1}{a^2} +\fr{8\pi
G_4}{3}\rho_{eECFT},~~~\rho_{eECFT}=-\fr{\tilde E_{SdS}}{V}.
 \eeq The equation (\ref{2ISR}) is well-defined even
at $a=r_{EH/CH}$. Thus this leads to $H_{SdS}^2=  1/\ell^2$ at
$a=r_{EH/CH}$. We note here that a spacelike brane can cross both
the event and cosmological horizons of the SdS black hole
background.

\subsection{MDW in TdS spaces}
Eq.(\ref{Tds}) together with $u^Mu^Ng_{MN}=1;~n^Mn^Ng_{MN}=-1$
leads to a spacelike brane \beq -h_{TdS}(r)dt^2 +
\fr{1}{h_{TdS}(r)} dr^2 ~~\to~~ -\Big(h_{TdS}(a) \dot t^2
-\fr{\dot a^2}{h_{TdS}(a)} \Big)d\tau^2= d\tau^2. \eeq Then the
induced line element is \beq ds^{2}_{4TdS}=d \tau^2 +a(\tau)^2
d\Sigma^2_k. \label{3INM} \eeq For a single TdS space, we
introduce the  brane tension $\sigma=3/(\kappa^2\ell)$ to obtain
the critical brane. The Israel junction condition of $\sqrt{ \dot
a^2-h_{TdS}(a)}=\fr{\kappa^2}{3}\sigma a $ leads to \beq
H^2_{TdS}=\fr{k}{a^2} +\fr{m_{TdS}}{a^4}. \label{3HHH} \eeq Here
we find a positive term of $m_{TdS}/a^4$ where $m_{TdS}$ is
defined at Eq.(\ref{3TQ}). This term behaves like a
radiation\cite{MYU2}. Further we have $m_{TdS}=16 \pi G_5M_{TdS}/3
V_3$ and $M_{TdS}=a \tilde E_{TdS}/\ell$. Then one finds a
CFT-radiation dominated universe \beq H^2_{TdS}= \fr{k}{a^2}
+\fr{8\pi G_4}{3}\rho_{ECFT},~~~\rho_{ECFT}=\fr{\tilde
E_{TdS}}{V}. \label{3CRA} \eeq The junction condition  holds even
at $a=r_{CH}$. Thus this leads to $H_{TdS}^2= 1/\ell^2$ at
$a=r_{CH}$.

As in the static the A(dS)/(E)CFT correspondences, we find the
positive energy density $\rho_{CFT}$ in the timelike brane moving
in the TAdS black hole background, while we find the negative
energy density $\rho_{eECFT}$ in the spacelike brane moving in the
SdS black hole background. Also one finds the positive energy
density $\rho_{ECFT}$ in the spacelike brane moving in the TdS
space background. This means that for the five-dimensional gravity
system with a cosmological constant, the dynamic A(dS)/(E)CFT
correspondences are consistent with the static A(dS)/(E)CFT
correspondences. The difference is that for the TAdS black hole we
obtain a CFT in the timelike brane and for the SdS black hole and
TdS space we have an ECFT in the spacelike brane. Especially, the
negative energy for the cosmological horizon of the SdS black hole
($E_4^{CSdS}$) persists in the energy density of $\rho_{eECFT}$.
Hence the cosmological horizon of the SdS black hole is still
problematic in view of the dS/ECFT correspondence.

It is found from Eqs.(\ref{HHH}), (\ref{2HHH}) and (\ref{3HHH})
that the Friedmann equation for the MDW in the TdS spaces can be
found from those of the TAdS black holes by replacing $k$ by $-k$.
Also we find that the Friedmann equation for the  MDW in the SdS
black hole is obtained from either the MDW in the HAdS black hole
or the MDW in the STdS space by substituting $m$ with $-m$. The
energy density term of $\rho_{(E)CFT}$ is included in TABLE  to
compare it with the sign of the boundary (E)CFT energy $E_4$. In
the case of the event horizon for  the SdS black hole, two are
different. This shows that the dS/ECFT correspondence for the SdS
black hole is not yet  established.

\section{Logarithmic corrections due to thermal fluctuations}

First we make corrections to the Bekenstein-Hawking entropy. The
corrected formula takes the form\cite{KM1,DMB,MP} \beq \label{CEN}
S=S_0-\fr{1}{2} \ln C_v + \cdots, \eeq where $C_v$ is the specific
heat of the given system at constant volume and $S_0$ denotes the
uncorrected Bekenstein-Hawking entropy. Here an important point is
that for Eq.(\ref{CEN}) to make sense, $C_v$ should be positive.
As is shown in TABLE, for the FAdS black hole and FTdS solution
one finds $C_v=3S_0$ without any approximation. However, other
cases (HAdS and SAdS black holes, CSdS, STdS and HTdS spaces) lead
to $C_v \simeq 3S_0$ when choosing large black holes
($r_{EH}^2>>\ell^2$) and large cosmological horizons
($r_{CH}^2>>\ell^2$). As far as $C_v \simeq 3S_0$ is guaranteed,
the logarithmic correction to the Bekenstein-Hawking entropy is
given by \beq \label{CENT} S^{TAdS,CSdS,TdS}=S_0-\fr{1}{2} \ln S_0
+ \cdots. \eeq Note that there is no correction to the event
horizon of the  SdS black hole (ESdS):
$S_{EH}^{ESdS}=S_{0}^{ESdS}$. Thus we do not consider this case
hereafter.  Logarithmic corrections to the Cardy-Verlinde formulae
are being performed by calculating the Casimir energy.  These are
found to be \beqa &&E_c^{TAdS}=k \fr{3\ell r_{EH}^2 V_3}{8
\pi G_5 R}+ \fr{3}{2} T \ln S_0,\\
 &&E_{c}^{CSdS}=-
\fr{3\ell
r_{CH}^2 V_3}{8 \pi G_5 R}+ \fr{3}{2} T \ln S_0,\\
 &&E_c^{TdS}=-k \fr{3\ell r_{CH}^2 V_3}{8 \pi G_5 R}+ \fr{3}{2} T
\ln S_0. \eeqa Substituting the above into the Cardy-Verlinde
formulae in Eqs.(\ref{cav1}), (\ref{cav2}) and (\ref{cav3}), one
finds \beqa && TAdS~ :~~\fr{ 2 \pi R}{3 \sqrt{|k|}}
\sqrt{|E_c(2E_4-E_c)|} \simeq S_0 + \fr{\pi R \ell T}{2kr_{EH}^3}
\Big( \fr{r_{EH}^4}{\ell^2}-kr_{EH}^2 \Big) \ln S_0, \\
&& CSdS~:~~\fr{ 2 \pi R}{3} \sqrt{|E_c(2E_4-E_c)|} \simeq S_0
-\fr{\pi R \ell T}{2r_{CH}^3}
\Big( \fr{r_{CH}^4}{\ell^2}+r_{CH}^2 \Big) \ln S_0, \\
&& TdS~:~~\frac{ 2 \pi R}{3 \sqrt{|k|}} \sqrt{|E_c(2E_4-E_c)|}
\simeq S_0 - \fr{\pi R \ell T}{2kr_{CH}^3} \Big(
\fr{r_{CH}^4}{\ell^2}+kr_{CH}^2 \Big) \ln S_0.  \eeqa All
coefficients in front of  $\ln S_0$ in the above relations are
transformed into the same expression as~\cite{myung} \beqa &&
\fr{\pi R\ell T}{2 k r_{EH}^3} \Big(
\fr{r^4_{EH}}{\ell^2} -kr_{EH}^2\Big)=\fr{(4E_4-E_c)(E_4-E_c)}{2(2E_4-E_c)E_c}, \\
&& -\fr{\pi R\ell T}{2  r_{CH}^3 }\Big(
\fr{r^4_{CH}}{\ell^2} +r_{CH}^2\Big)=\fr{(4E_4-E_c)(E_4-E_c)}{2(2E_4-E_c)E_c}, \\
&& -\fr{\pi R\ell T}{2 k r_{CH}^3 }\Big( \fr{r^4_{CH}}{\ell^2}
+kr_{CH}^2\Big)=\fr{(4E_4-E_c)(E_4-E_c)}{2(2E_4-E_c)E_c}. \eeqa
Finally we obtain the same form  of corrections to the
Cardy-Verlinde formula as \beq \label{CCV}S^{TAdS,CSdS,TdS}_{CV}
\simeq \fr{ 2 \pi R}{3 \sqrt{|k|}} \sqrt{|E_c(2E_4-E_c)|}-
\fr{(4E_4-3E_c)E_4}{2(2E_4-E_c)E_c} \ln \Big(\fr{ 2 \pi R}{3
\sqrt{|k|}} \sqrt{|E_c(2E_4-E_c)|}\Big)\eeq where we recover the
Cardy-Verlinde formula for the cosmological horizon of the SdS
black hole when $k=1$. The Cardy-Verlinde formula expresses the
holography which means that the bulk thermal properties can be
determined from their boundary thermal CFT. Note that after
logarithmic corrections to the Cardy-Verlinde formula due to
thermal fluctuation, they take still the same form. This means
that there is no  crucial difference between the event horizon of
the TAdS black hole and the cosmological horizons of the TdS
spaces and the SdS black holes.
\begin{figure}
\begin{center}
\epsfig{file=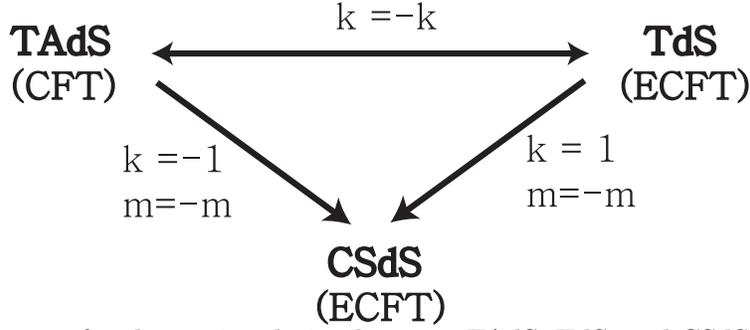,width=0.6\textwidth} \caption{The diagram
for the static relation between TAdS, TdS, and CSdS. The CFT in
TAdS black hole is defined on the timelike boundary, while the
ECFT in TdS spaces and CSdS black hole is defined on the spacelike
boundary.}
\end{center}
\end{figure}
Fig.1 shows a  close relationship between TAdS, TdS, and CSdS
after logarithmic corrections.

\section{Logarithmic corrections to the MDW}
In section IV, the brane cosmology has been studied in the
framework of the dynamic A(dS)/(E)CFT correspondences. The brane
starts with (big bang) inside the event horizon (cosmological
horizon), crosses the horizon, and expands until it reaches
maximum size. And then the brane contracts,
 it falls the horizon again and finally disappears (big crunch). An
observer in the bulk space finds two interesting moments  when the
brane crosses the past (future)  horizon. Authors in ref.\cite{SV}
showed  that at these times the Friedmann equation controlling the
cosmological expansion (contraction) coincides with a
Cardy-Verlinde formula for the entropy of the CFT on the brane.
That is, the location of the event horizon (cosmological horizon)
is a holographic point. On the Cardy-Verlinde formula side, one
can obtain its logarithmic correction due to thermal fluctuations
of the bulk gravity system. However, it is not easy to obtain its
corresponding term in the Friedmann equation. Up to now we don't
know how to embed the logarithmic term into the Friedmann
equation.  The only way is to use a holographic point if one
assumes that the Cardy-Verlinde formula coincides the Friedmann
equation when the MDW crosses the  horizon of the bulk spacetime.
Using the Hubble entropy of $ H^2=\Big( \fr{2G_4}{V}
\Big)S_{CV}^2$ and Eq.(\ref{CCV}), one finds the modified
relations at the holographic point \beqa \label{holo} &&
H^2_{TAdS}=\fr{1}{\ell^2}-\fr{2G_4}{V\ell} \ln S_0
=-\fr{k}{a_{EH}^2} +\fr{8\pi G_4}{3}\rho_{CFT} -\fr{2G_4}{V\ell}
\ln S_0, \\
&& H^2_{CSdS}=\fr{1}{\ell^2}-\fr{2G_4}{V\ell} \ln S_0
=\fr{1}{a_{CH}^2} +\fr{8\pi G_4}{3}\rho_{eECFT} -\fr{2G_4}{V\ell}
\ln S_0, \\
&& H^2_{TdS}=\fr{1}{\ell^2}-\fr{2G_4}{V\ell} \ln S_0
=\fr{k}{a_{CH}^2} +\fr{8\pi G_4}{3}\rho_{ECFT} -\fr{2G_4}{V\ell}
\ln S_0. \eeqa The expansion (contraction) rate at a holographic
point decreases in comparison  with  the case without correction :
$H^2_{TAdS,SdS,TdS}=1/\ell^2$. We assume to extend these relations
to the modified Friedmann equations
 \beqa
\label{modf1} && H^2_{TAdS}=-\fr{k}{a^2} +\fr{8\pi
G_4}{3}\rho_{CFT} -\fr{2G_4}{V\ell}
\ln S_0, \\
\label{modf2}&& H^2_{CSdS}=\fr{1}{a^2} +\fr{8\pi
G_4}{3}\rho_{eECFT} -\fr{2G_4}{V\ell}
\ln S_0, \\
\label{modf3}&& H^2_{TdS}=\fr{k}{a^2} +\fr{8\pi G_4}{3}\rho_{ECFT}
-\fr{2G_4}{V\ell} \ln S_0 \eeqa which are valid for  other points
in the bulk background. Here $V=a^3V_3$ and $S_0=V_3 a^3/4G_5$. In
the absence of logarithmic corrections, the above equations reduce
to Eqs.(\ref{1CRA}), (\ref{2CRA}) and (\ref{3CRA}) respectively.
Let us transform the modified Friedmann equation to the
conservation of energy for a point particle moving under the
one-dimensional potential $V$ as\beqa \label{ecl} && {\dot
a}^2=-k-V_{TAdS},~~V_{TAdS}= -\fr{m_{TAdS}}{a^2}
+\fr{2G_4}{V_3\ell a}
\ln \Big( \fr{V_3a^3}{2\ell G_4}\Big), \\
&& {\dot a}^2=1-V_{CSdS},~~V_{CSdS}= \fr{m_{SdS}}{a^2}
+\fr{2G_4}{V_3\ell a}
\ln \Big( \fr{V_3a^3}{2\ell G_4}\Big), \\
&& {\dot a}^2=k-V_{TdS},~~V_{TdS}= -\fr{m_{TdS}}{a^2}
+\fr{2G_4}{V_3\ell a} \ln \Big( \fr{V_3a^3}{2\ell G_4} \Big).
\eeqa
\begin{figure}
\begin{center}
\epsfig{file=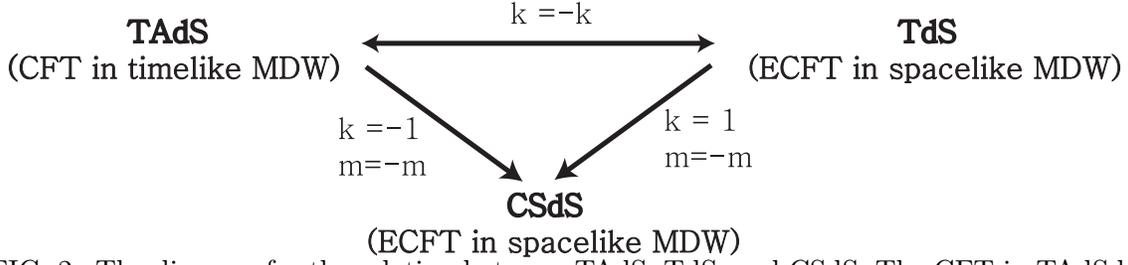,width=0.9\textwidth} \caption{The diagram
for the relation between TAdS, TdS, and CSdS. The CFT in TAdS
black hole is defined on the timelike MDW, while the ECFT in TdS
spaces and CSdS black hole is defined on the spacelike MDW.}
\end{center}
\end{figure}
Fig.2 shows a  close relationship  between TAdS, TdS, and CSdS
after logarithmic corrections to the Friedmann equation. For the
cosmological implication of  logarithmic corrections, we consult
ref.\cite{LNOO} for the TAdS and TdS cases and ref.\cite{NOO} for
the SdS case. The logarithmic correction ($\ln a/a$) to the
timelike (spacelike) MDW under the HAdS black hole (STdS space)
give us  additional bouncing cosmologies without any big bang
singularity, in compared with others of the big bang $\to$ big
crunch. Although there exists logarithmic correction to the
spacelike MDW in the SdS space, the physical interpretation is
still unclear because of the presence of the exotic energy density
\footnote{However, for the timelike MDW which crosses only the
event horizon of  the SdS black hole, one finds the positive
energy density\cite{CM}.}:$\rho_{eECFT}=-\tilde E/V$. Finally we
are not sure that these logarithmic corrections could be included
after corrections to the Friedmann equations.

\section{ Discussion}
We show that all thermal properties of the TdS spaces can be found
from those of the TAdS black holes by replacing $k$ by $-k$. This
means that the cosmological horizon in TdS spaces is nearly
identical  with  the event horizon of the TAdS black holes.  Also
we find that all thermal information for the cosmological horizon
of the SdS black hole is obtained from either the hyperbolic-AdS
black hole or the Schwarzschild-TdS space by substituting $m$ with
$-m$. This means  that the cosmological horizon of the SdS black
hole is nothing special.

Concerning the SdS black hole with the cosmological and event
horizons, the dS/CFT correspondence is problematic because as is
shown in TABLE, the event horizon has an exotic (negative) energy
density in the Friedmann equation while the cosmological horizon
has a negative CFT energy and an exotic energy density. Also the
negative Casimir energy for the cosmological horizon implies the
non-unitary CFT. In addition the event horizon inside the
cosmological horizon has the negative specific heat, which means
that it is thermodynamically unstable. Hence the SdS black hole
does not seems to be a toy model to study the cosmological horizon
in asymptotically de Sitter space because it has two
horizons\footnote{The other thermal approach to the
four-dimensional SdS black hole, see ref.\cite{SHAN}.}.

We suggest that the TdS spaces including a cosmological
singularity are better candidates for studying thermal property
for the cosmological horizon than the SdS black hole. Considering
the close relationship between the TdS spaces and the TAdS black
holes, the cosmological horizon is nothing but the event horizon.
Also the dS/ECFT correspondence is well established for the TdS
spaces in conjunction with the AdS/CFT correspondence for the TAdS
black holes.  However, in the wave equation approach to find the
absorption cross section, two may be different because their
working spaces are different (the TdS case is compact whereas the
TAdS is non-compact)\cite{myung1}.

\section*{Acknowledgments}

This work was supported in part by KOSEF, Project No.
R02-2002-000-00028-0.


\begin{thebibliography}{99}
\bibitem{Per} S. Perlmutter et al.(Supernova Cosmology Project),
Astrophys. J. {\bf 483}, 565(1997)[astro-ph/9608192].


\bibitem{CDS}R. R. Caldwell, R. Dave, and  P. J. Steinhard, Phys. Rev. Lett.
{\bf 80}, 1582(1998)[astro-ph/9708069].

\bibitem{Gar}P. M. Garnavich et al.,  Astrophys. J. {\bf 509},
74(1998)[astro-ph/9806396].

\bibitem{Wit} E. Witten,
hep-th/0106109.

\bibitem{HKS}S. Hellerman, N. Kaloper, and L. Susskind, JHEP {\bf 0106},
003(2001)[hep-th/0104180].

\bibitem{FKMP}  W. Fischler, A. Kashani-Poor,
R. McNees, and  S. Paban, JHEP {\bf 0107}, 003
(2001)[hep-th/0104181].

\bibitem{BOU} R. Bousso, JHEP {\bf 0011}, 038 (2000)[hep-th/0010252];
 R. Bousso, JHEP {\bf 0104}, 035 (2001)[hep-th/0012052];

\bibitem{TDS} V. Balasubramanian, J. de Boer, and D. Minic, Phys. Rev.  {\bf
D65} (2002) 123508 [hep-th/0110108];
R. G. Cai, Y. S. Myung, and Y. Z. Zhang, Phys. Rev. {\bf D65}
(2002) 084019 [hep-th/0110234];
 Y. S. Myung, Mod. Phys. Lett.{\bf A16} (2001) 2353 [hep-th/0110123];
 A. M. Ghezelbach and R. B. Mann, JHEP {\bf 0201} (2002) 005
  [hep-th/0111217].

\bibitem{STR} A. Strominger, JHEP {\bf 0110}, 034
(2001)[hep-th/0106113].

\bibitem{GHaw} G. W. Gibbons and S. W. Hawking, Phys. Rev.
{\bf D15}, 2738(1977).

\bibitem{DMB} S. Das, P. Majumdar, and R. K. Bhaduri,
Class. Quant. Grav. {\bf 19} (2002) 2355 [hep-th/0111001].

\bibitem{KM1} R. K. Kaul and P. Majumdar, Phys. Lett. {\bf B439}
(1998) 267 [gr-qc/9801080];

 R. K. Kaul and P. Majumdar, Phy. Rev. Lett, {\bf 56}
(2000)5255 [gr-qc/0002040];

 S. Carlip, Class. Quant. Grav. {\bf 17} (2000) 4175
 [gr-qc/0005017];

 T. R. Govindarajan, R.K. Kaul and V. Suneeta,
Class. Quant. Grav. {\bf 18} (2001) [gr-qc/0104010];

 D. Birmingham and S. Sen, Phys. Rev. {\bf D63} (2001) 047501
[hep-th/0008051];

J. Jing and Mu-Lin Yan, Phys. Rev. {\bf D63} (2001) 024003
[gr-qc/0005105].



\bibitem{BIR} D. Birmingham, Class.  Quant. Grav. 16 (1999) 1197 [hep-th/9808032].

\bibitem{MP} S. Mukherji and S. S. Pal, JHEP {\bf 0205} (2002) 026
 [hep-th/0205164].

\bibitem{CAI1} R. G. Cai, Phys. Rev. {\bf D63} (2001) 124018
  [hep-th/0102113].

\bibitem{LNOO} J. E. Lidsey, S. Nojiri, S. Odintsov, and S .
Ogushi, Phys. Lett. {\bf B544} (2002) 337
  [hep-th/0207009].

\bibitem{CAI2}R. G. Cai, Phys. Lett. {\bf B525} (2002)
331 [hep-th/0111093].

\bibitem{CM} R. G. Cai and Y. S. Myung, Phys. Rev. {\bf D67} (2003) 124021
  [hep-th/0210272].

\bibitem{NOO} S. Nojiri, S. Odintsov, and S. Ogushi, Int. J. Mod. Phys. {\bf A18} (2003) 3395
[hep-th/0212047];

\bibitem{myung} Y. S. Myung, hep-th/0308191, to appear in Physics Letters B.

\bibitem{HOL} J. Maldacena, Adv. Theor. Math. Phys. {\bf 2}
(1998) 231 [hep-th/9711200];
 S.S Gubser, I.R. Klebanov, and A.M. Polyakov,
  Phys. Lett. {\bf B428} (1998) 105 [hep-th/9802109];
E. Witten, Adv. Theor. Math. Phys. {\bf 2} (1998) 253
 [hep-th/9802150].

\bibitem{VER} E. Verlinde, hep-th/0008140.

\bibitem{witten} E. Witten,
Adv. Theor. Math. Phys. {\bf 2} (1998) 505 [ hep-th/9803131].
\bibitem{SET}  M. R. Setare, hep-th/0308106.


\bibitem{SV} I. Savonije and E. Verlinde, Phys. Lett. B507 (2001) 305 [hep-th/0102042].

\bibitem{MYU2} Y. S. Myung, Phys. Lett. {\bf B531} (2002)1
 [hep-th/0112140].
\bibitem{SHAN} S. Shankaranarayanan, Phys. Rev. {\bf D67} (2003) 084026
[gr-qc/0301090].

\bibitem{myung1} Y. S. Myung, hep-th/0304231.

\end{thebibliography}
\end{document}